\documentclass[conference]{IEEEtran}
\IEEEoverridecommandlockouts
\usepackage{multicol}
\usepackage{multirow}
\usepackage{cite}
\usepackage[numbers]{natbib}
\usepackage{amsmath,amssymb,amsfonts}
\usepackage{algorithmic}
\usepackage{authblk}
\usepackage{graphicx}
\usepackage{textcomp}
\usepackage{xcolor}
\usepackage{stfloats}
\usepackage{colortbl}

\def\BibTeX{{\rm B\kern-.05em{\sc i\kern-.025em b}\kern-.08em
    T\kern-.1667em\lower.7ex\hbox{E}\kern-.125emX}}

\title{Multiple Consistency-guided Test-Time Adaptation for Contrastive Audio-Language Models with Unlabeled Audio\\
}
\author{
    \textit{Gongyu Chen$^2$$^*$, Haomin Zhang$^1$$^*$, Chaofan Ding$^1$, Zihao Chen$^1$, Xinhan Di$^1$} \\
    $^1$AI Lab, Giant Network.\\
    $^2$College of Computer and Science, Zhejiang University \\
    \small\texttt{22221269@zju.edu.cn, \{zhanghaomin, dingchaofan, chenzihao, dixinhan\}@ztgame.com}
}
\begin{document}

\maketitle

\begin{abstract}
One fascinating aspect of pre-trained Audio-Language Models (ALMs) learning is their impressive zero-shot generalization capability and test-time adaptation (TTA) methods aiming to improve domain performance without annotations. However, previous test time adaptation (TTA) methods for ALMs in zero-shot classification tend to be stuck in incorrect model predictions. In order to further boost the performance, we propose multiple guidance on prompt learning without annotated labels. First, guidance of consistency on both context tokens and domain tokens of ALMs is set. Second, guidance of both consistency across multiple augmented views of each single test sample and contrastive learning across different test samples is set. Third, we propose a corresponding end-end learning framework for the proposed test-time adaptation method without annotated labels. We extensively evaluate our approach on $12$ downstream tasks across domains, our proposed adaptation method leads to $4.41\%$ (max $7.50\%$) average zero-shot performance improvement in comparison with the state-of-the-art models. 
\end{abstract}


\section{Introduction} 
In recent years, audio-language models (ALMs) learning under language supervision exhibit promising zero-shot transferability \citep{elizalde2024clap2, deshmukh2024pengi, gong2024listenthinkunderstand, wu2024laion}. Pretrained on large-scale paired audio-text datasets, ALMs like CLAP \citep{clap1, elizalde2024clap2} can generalize well to new audio data without requiring task-specific fine-tuning. However, its zero-shot performance heavily relies on the effectiveness of prompts used \citep{clap1} during inference. The process of manually designing these prompts is both labor-intensive and computationally demanding. Furthermore, while few-shot learning \citep{li2023fewshotprompt, liang2023fewshotadapter} offers an alternative by leveraging limited labeled data, it is not always feasible, especially when dealing with out-of-domain (OOD) scenarios, where model performance tends to degrade significantly.

Adaptation methods have been extensively explored to fine-tune learnable prompts using a limited number of labeled test samples, improving model performance in zero-shot scenarios. Context optimization (CoOp) is developed to optimize context-aware prompts with minimal labeled data \cite{coop}, it is then extended by proposing conditional context optimization (CoCoOp) \cite{cocoop}, which dynamically adjusts prompts based on input instances, improving generalization across both seen and unseen classes. Then, self-regulating prompts \cite{promptsrc} explicitly steer the prompts to learn a representation space by regularization that maximize performance on downstream tasks without compromising CLIP \citep{clip} generalization. In addition, generalization of large foundation models is improved through enforcing consistency \cite{roy2024consistencyguidedPLVL} on two perturbed inputs and combining two dominant paradigms of tuning, prompting and adapter. However, these existing methods require access to annotated target data which is expensive and impractical for deployment.

Besides, several unsupervised methods have been proposed to improve model performance under distribution shifts. Test-Time Training (TTT) \cite{TTT} introduces self-supervised learning at test time by updating model parameters based on each test sample, which has shown improvements in generalization. Extensions like TTT++ \cite{ttt++} and Tent \cite{Wang2021TentFT} further optimize test-time adaptation by minimizing entropy and exploring contrastive learning objectives. Moreover, masked autoencoders have been applied to this domain, as shown in \cite{TTTmaskae}, enhancing generalization by learning from partially visible inputs. Another approach, MEMO \cite{MEMO}, augments and adapts models at test time to improve robustness against distribution shifts. Test-Time Prompt Tuning (TPT) \cite{TPT, deshmukh2024DA} tries optimizing the learnable prefix tokens by minimizing the entropy. However, these existing unsupervised methods tend to be stuck in incorrect model predictions in test-time adaptation, and the performance is beyond satisfaction.

Therefore, in order to boost the performance in the test-time adaptation without annotated audio samples, we propose an end-end framework based on multiple type of guidance. First, guidance of consistency on both context tokens and domain tokens of audio-language models (ALMs) is set in the test time adaptation. Second, guidance of both consistency across multiple augmented views of each single test sample and contrastive learning across different test samples is set. Third, a corresponding end-end learning framework for the proposed test-time adaptation method is built without annotated labels. In the evaluation, extensive experiments are conducted on $12$ downstream tasks across domains. In the comparison with the state-of-the-art models, the proposed adaptation method leads to $4.41\%$ (max $7.50\%$) average zero-shot performance.   

 

\begin{figure*}[t]
\centering
\includegraphics[width=\textwidth]{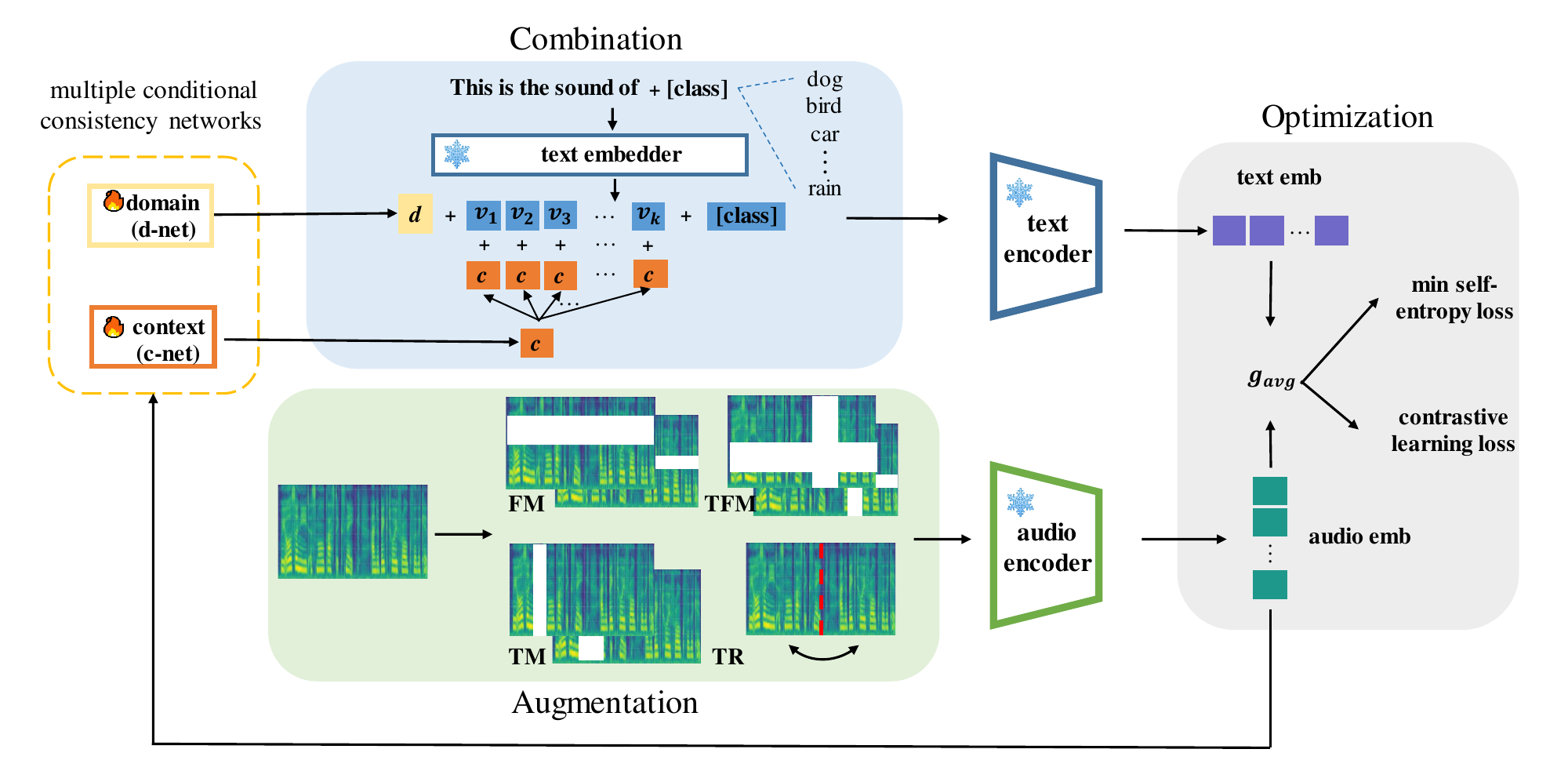}  
\caption{Our end-end Test-time Tuning framework. (1) \textbf{Augmentation.} Four augmentations are performed on the raw audio, with Time Reorder first cutting the spectrum in half and then swapping the order before and after. (2) \textbf{Combination.} Multiple conditional consistency networks receive the audio embedding and generate learnable tokens combined with the original prompt in two ways. (3) \textbf{Optimization.} The minimum self-entropy loss and contrastive learning loss are calculated using the average distribution.}
\label{fig:pdf}
\end{figure*}

\section{Background} 
We regard test-time prompt tuning as a way to provide the model with the tuned representation tailored to the test samples, which helps precisely retrieve the knowledge of CLAP. At the inference stage, the only information available is each test sample \(X_{test}\) without label information. Therefore, manages to optimize the prompt \(p\) at test time based on each test sample. In general, our objective can be formulated in the form of
\[
p^* = \arg\min_{p} \mathcal{L}(\mathcal{F}, p, X_{test})
\]
\noindent where \(\mathcal{F}=\{E_{audio}, E_{text}\}\) is a CLAP model with \(E_{audio}\) and \(E_{text}\) being the audio and text encoders, \(\mathcal{L}\) denotes some carefully constructed loss.

Specifically, the learnable prompt $p$ is perturbed into two components, $p_1$ and $p_2$, which are the domain-aware prompts and context-aware prompts. Domain-aware prompts ($p_1$) are applied by pretending a learnable sequence to the original text input, introducing a domain-specific perturbation \cite{deshmukh2024DA} that adjusts the model's interpretation based on the characteristics of the input data without altering the core prompt. On the other hand, context-aware prompts ($p_2$) involve replicating and appending learnable tokens to the text input, perturbing the core prompt and injecting additional context \cite{cocoop,coop}. These combined perturbations help the model to adapt to different tasks (domains).  
\[
p_1^*,p_2^* = \arg\min_{p_1, p_2} \mathcal{L}(\mathcal{F}, concat(p_1,p_2+p_{origin}), X_{test})
\]

\section{Multiple Consistency-guided Test-Time Adaptation} 

Our approach is divided into three main components: conditional context-aware consistency adaptation, conditional domain-aware consistency adaptation, and corresponding end-end test-time tuning adaptation framework. Below, we explain each in detail.

\subsection{Conditional Context-Aware Consistency Adaptation}

In this step, we employ a CLAP model pre-trained with a speech encoder and a text encoder. Given an audio sample, we first apply specaugment \cite{deshmukh2024DA} to generate multiple augmented views of the audio. These augmented audio samples are then passed through the CLAP's speech encoder to extract the audio features $v_i$. Simultaneously, we introduce a learnable context-aware prompt $p_{\text{context}}$, which is generated conditionally using a lightweight neural network with learnable parameters $\theta_{1}$. This prompt is combined with the text embeddings produced by the CLAP text encoder to form the context text features $\mathbf{t}_{context}$:
\[
\mathbf{t}_{context} = f_{text}(p_{context}(\theta_{1},v_i) + p_{origin})
\]
where $f_{\text{text}}$ is the CLAP text encoder and $p_{\text{context}}$ is the context-aware prompt. The generated context text features and audio features are then used to calculate a joint distribution for the test-time adaptation. 

\subsection{Conditional Domain-Aware Consistency Adaptation}

\begin{table*}[htbp]
\centering
\caption{Zero-shot and Test-time adaptation performance on 12 downstream tasks.}
\label{mainTabel}
\renewcommand{\arraystretch}{1.15}
\begin{tabular}{l| c | c c c c c c c c c c c c }
\hline
Dataset & Avg.↑ & ESC50 & US8K & D17T4 & SESA & GT MS & Genre & Opera & CREM & RAVD & Vocal & TU17 & NSyn  \\
\hline
Zero-Shot (CLAP \citep{elizalde2024clap2})  & 62.93 & 93.90 & 82.30 & 46.60 & 64.95 & 99.20 & 58.40 & 46.60 & 30.00 & 31.54 & 79.97 & 53.80 & 68.00  \\ \hline
DA CLAP (one \citep{deshmukh2024DA})  & 64.94 & 93.35 & 85.26 & 50.96 & 73.30 & 99.21 & 61.00 & 47.45 & 29.92 & \textbf{34.25} & 82.14 & 54.19 & 68.28  \\
DA CLAP (five \citep{deshmukh2024DA}) & 65.92 & \textbf{95.05} & 85.21 & 52.30 & 74.35 & 100.0  & 63.20 & 50.42 & 31.27 & 33.19 & 82.40 & 54.38 & 69.23  \\  
DA CLAP (whole \citep{deshmukh2024DA})  & 64.57 & 94.15 & 85.83 & 51.33 & 69.52 & 100.0  & 58.06 & 44.49 & \textbf{37.07} & 32.26 & 80.79 & 57.22 & 64.09  \\  \hline
ours   & \textbf{68.83} & 94.65 & \textbf{86.60} & 51.69 & \textbf{76.19} & 100.0  & \textbf{65.87} & 68.64 & 34.29 & 33.77 & 82.21 & 58.46 & \textbf{73.54}  \\  
-- w/o context-aware        & 67.51 & 94.55 & 84.67 & 50.79 & 75.24 & 100.0  & 61.86 & \textbf{70.34} & 32.56 & 30.06 & 81.37 & \textbf{59.51} & 69.19  \\
-- w/o domain-aware        & 66.96 & 94.15 & 82.99 & \textbf{54.29} & 73.33 & 100.0  & 64.67 & 57.63 & 34.13 & 32.91 & \textbf{82.40} & 57.59 & 69.39  \\
\hline

\hline
\end{tabular}
\end{table*}

In addition to the context-aware adaptation, we also generate domain-aware prompts $p_{\text{domain}}$, which are designed to capture domain-specific characteristics of the audio samples. These domain prompts are prepended to the text sequence, resulting in domain-conditioned text features $\mathbf{t}_{\text{domain}}$:
\[
\mathbf{t}_{domain} = f_{text}(p_{domain(\theta_{2},v_i)} \oplus p_{origin})
\]
where $f_{\text{text}}$ is the CLAP text encoder and $p_{\text{domain}}$ is the domain-aware prompt. The generated domain text features and audio features are then used to calculate a joint distribution for the test-time adaptation. Similarly, the domain-aware learnable prompts are generated through a lightweight neural network with learnable parameters $\theta_{2}$, conditioned on the audio features extracted from the augmented views. 

\subsection{End-end Test-time Tuning}

In order to apply the proposed two types of consistency for the test-time adaptation of audio-language models (ALMs) without labels, we propose an end-end framework to boost the performance of ALMs (Figure \ref{fig:pdf}). 

In detail, the end-end framework consists of two lightweight networks (c-net and d-net), an audio-language model with two parts(audio encoder and text encoder), a text embedder and an augmenter. In order to train the end-end framework, three phases including the augmentation, combination and optimization are represented as the following.

\subsubsection{Augmentation and Combination}
The audio file is converted to a mel spectrogram (\(x \in \mathbb{R}^{T\times F}\)). Rather than directly passing \(x\) to the audio encoder, it is first augmented using multiple random yet controlled augmentations. The augmentation is masking-based and inspired by SpecAug \citep{Park2019SpecAugmentAS}:
\[
\hat{\mathcal{X}} = [x, \text{TM}(x), \text{FM}(x), \text{TFM}(x), \text{TR}(x)]
\]
where TM, FM, TFM and TR denote \textit{Time Masking}, \textit{Frequency Masking} \textit{Time \& Frequency Masking} and \textit{Time Reorder}, respectively \cite{deshmukh2024DA}. The augmented audio embedding \( \hat{v} \in \mathbb{R}^{M \times d} \) and learnable text embeddings \( \hat{u} \in \mathbb{R}^{N \times d} \) are used to compute a dot product:
\[
g = \hat{v} \cdot \hat{u}^T
\]
\[
\hat{v} = f_{audio}(\hat{x}), \hat{x}\in \mathcal{\hat{X}}
\]
\[
\hat{u} = f_{text}(p_{context}(\theta_{1},\hat{v}) + p_{origin} \oplus p_{domain}(\theta_{2},\hat{v}))
\]
where \( g \in \mathbb{R}^{M \times N} \). We use softmax to convert \( g \) into a valid probability distribution across classes \( N \). After softmax, \( g \) is the probability of each class for a total of \( M \) augmented views of the audio. The probability distribution \( g_{avg} \) is then averaged along the different augmentations \( M \) as $g_{avg} = \frac{1}{M} \sum_{i=1}^{M} g_{i}$.

\subsubsection{Optimization}
In order to train the proposed end-end framework (Figure \ref{fig:pdf}), two types of losses are applied, the consistency loss \cite{deshmukh2024DA} and a contrastive learning loss. First, the consistency loss $\mathcal{L}_{\text{consistency}}$ calculates self-entropy from the distributions derived from the audio and text features. 
\[
\mathcal{L}_{consistency} = - g_{avg} \log g_{avg}
\]
where $g_{avg}$ is the average probability distribution along the different \( M \) augmentations.

Besides, a contrastive learning loss \citep{Yamashita_2024_CVPR} is applied to encourage diversity among the classification distributions of different audio samples between the distributions of any two different audio(wav) samples $k1$ and $k2$:
\[
\mathcal{L}_{contrastive} = -\sum_{k1 \neq k2} \text{MSE}(g_{k1}, g_{k2})
\]

The final loss $\mathcal{L}_{\text{final}}$ is a weighted sum of the minimal entropy loss and the contrastive learning loss:
\[
\mathcal{L}_{final} = \mathcal{L}_{consistency} + \lambda_{contrastive} \mathcal{L}_{contrastive}
\]
where $\lambda_{\text{contrastive}}$ is a hyperparameter that controls the balance between the two losses.
\begin{table*}[htbp]
\centering
\caption{Cross-domain generalization (ours/DA CLAP \citep{deshmukh2024DA}) on 12 downstream tasks.}
\label{crossTable}
\setlength{\tabcolsep}{2.5pt}
\renewcommand{\arraystretch}{1.2}
\begin{tabular}{l| c | c c c c c c c c c c c c |c}
\hline
Dataset & ZS & ESC50 & US8K & D17T4 & SESA & GT MS & Genre & Opera & CREM & RAVD & Vocal & TU17 & NSyn & Avg. \\
\hline

ESC50 & 93.9 & \cellcolor{blue!30}94.7/93.4 & 82.9/81.8 & 48.0/46.7 & 72.4/72.2 & 99.2/100.0 & 57.6/60.6 & 60.2/40.6 & 34.4/26.2 & 34.4/31.2 & 81.3/81.2 & 57.5/54.3 & 67.2/67.5 & \textbf{65.8}/63.0 \\
US8K  & 82.3 & 94.2/90.8 & \cellcolor{blue!30}86.6/85.3 & 45.8/49.9 & 73.3/67.4 & 99.2/97.7 & 56.0/54.4 & 48.3/44.1 & 35.3/23.6 & 33.2/32.5 & 80.7/76.7 & 58.5/45.1 & 64.0/62.7 & \textbf{64.6}/60.8 \\
D17T4 & 46.6 & 95.1/90.6 & 86.3/84.7 & \cellcolor{blue!30}51.7/51.0 & 69.5/69.4 & 99.2/96.9 & 58.2/54.7 & 55.9/44.9 & 31.0/26.9 & 25.7/27.0 & 82.4/76.0 & 56.8/45.5 & 66.1/62.8 & \textbf{64.8}/60.9 \\
SESA  & 65.0 & 94.5/92.4 & 84.3/81.5 & 47.7/45.3 & \cellcolor{blue!30}76.2/73.3 & 99.2/99.2 & 57.8/56.3 & 54.7/33.5 & 34.5/28.0 & 33.4/32.2 & 80.8/81.9 & 58.4/51.6 & 66.5/65.3 & \textbf{65.7}/61.7 \\
GT MS & 99.2 & 94.5/90.8 & 84.4/85.0 & 49.6/49.7 & 72.4/65.6 & \cellcolor{blue!30}100.0/99.2 & 57.3/54.7 & 56.8/36.0 & 27.3/25.0 & 30.8/27.6 & 80.8/75.9 & 57.3/45.3 & 65.1/60.4 & \textbf{64.7}/59.6 \\
Genre & 58.4 & 94.3/92.6 & 84.3/82.4 & 48.9/46.2 & 72.4/72.6 & 98.4/98.4 & \cellcolor{blue!30}65.9/61.0 & 57.6/36.4 & 25.1/29.5 & 30.0/32.1 & 81.4/82.2 & 56.8/49.8 & 65.9/64.1 & \textbf{65.1}/62.3 \\
Opera & 46.6 & 94.3/92.4 & 84.4/82.6 & 49.0/46.7 & 69.5/73.0 & 99.2/99.2 & 57.5/56.6 & \cellcolor{blue!30}68.6/47.5 & 25.9/27.6 & 32.7/32.3 & 81.4/82.5 & 56.6/53.6 & 65.3/64.5 & \textbf{65.4}/63.2 \\
CREM  & 30.0 & 94.4/92.6 & 84.1/81.9 & 48.2/46.1 & 72.4/73.6 & 99.2/99.2 & 57.8/56.7 & 55.5/36.0 & \cellcolor{blue!30}34.3/29.9 & 34.2/32.4 & 81.6/82.2 & 57.5/52.6 & 66.5/65.0 & \textbf{65.5}/62.4 \\
RAVD  & 31.5 & 94.2/92.5 & 84.0/81.4 & 48.1/45.4 & 72.4/74.5 & 99.2/99.2 & 58.2/59.1 & 58.9/37.7 & 32.9/29.4 & \cellcolor{blue!30}33.8/34.3 & 82.2/82.0 & 57.6/52.2 & 66.0/65.2 & \textbf{65.6}/62.7 \\
Vocal & 80.0 & 94.4/92.8 & 84.2/81.9 & 49.1/46.1 & 69.5/74.1 & 99.2/99.2 & 57.3/58.2 & 58.1/36.4 & 27.1/28.3 & 31.4/32.2 & \cellcolor{blue!30}82.2/82.2 & 57.1/51.9 & 66.3/65.0 & \textbf{64.7}/62.4 \\
TU17  & 53.8 & 94.4/92.4 & 84.1/81.4 & 48.0/45.3 & 72.4/73.9 & 99.2/99.2 & 58.0/56.4 & 55.5/33.5 & 30.5/28.0 & 32.1/32.3 & 81.5/81.9 & \cellcolor{blue!30}58.5/54.2 & 66.0/65.2 & \textbf{65.0}/62.0 \\
NSyn  & 68.0 & 93.2/92.5 & 81.8/81.2 & 48.9/45.6 & 73.3/73.6 & 99.2/99.2 & 57.9/55.8 & 57.2/34.3 & 32.1/27.3 & 31.7/32.2 & 81.6/81.5 & 56.5/51.4 & \cellcolor{blue!30}73.5/68.3 & \textbf{65.6}/61.9 \\
\hline
Avg.  & 62.9 & \textbf{94.3}/92.1 & \textbf{84.3}/82.6 & \textbf{48.6}/47.0 & \textbf{72.1}/71.9 & \textbf{99.2}/98.9 & \textbf{58.3}/57.0 & \textbf{57.3}/38.4 & \textbf{30.9}/27.5 & \textbf{31.9}/31.5 & \textbf{81.5}/80.5 & \textbf{57.4}/50.6 & \textbf{66.5}/64.7 & \textbf{65.2}/61.9 \\
\hline
\end{tabular}
\end{table*}

\section{Results} 

\subsection{Experimental Results}

\subsubsection{Training details}

The contrastive audio language model \citep{elizalde2024clap2} is used for our experiments \citep{deshmukh2024DA}. Besides, the audio encoder is HTSAT \citep{HTSAT} and the text encoder is a modified GPT2 \citep{deshmukh2023TACMwithoutAudio, Radford2019GPT2}. The audio is sampled at 44.1 kHz and converted to log Mel spectrograms with 64 Mel bins, a hop size of 320, and a window size of 1024 in the range of 50-8000 Hz range. Our multiple conditional consistency networks are all MLP with three layers. We augment each raw audio 50 times to capture multiple views. The optimizer used is AdamW \citep{adamw} with a learning rate of 1e-6. 
\subsubsection{Evaluation} 

We benchmark our method on $12$ downstream tasks \citep{turian2022hear} across the domains of Sound Event Classification, Acoustic Scene Classification, Vocal Sounds, Music, Surveillance, and Speech Emotion Recognition. The datasets used are: ESC50 \citep{Piczak2015ESC50}, UrbanSound8K \citep{US8K}, DCASE2017 Task4 \citep{DCASE2017}, TUT 2017, GTZAN Music Speech \citep{Tzanetakis2001GTZAN}, GTZAN Genres \citep{Tzanetakis2001GTZAN}, Beijing Opera Percussions \citep{turian2022hear}, CREMA-D \citep{CREMAD}, RAVDESS \citep{RAVDESS}, Vocal Sound \citep{Gong2022VocalsoundAD}, SESA \citep{SESA}, NS Instrument family \citep{ns_inst}. The datasets have varied audio duration, classes, files, and setups. For example, the audio duration ranges from 3 to \(\geq\) 35 seconds, classes range from binary to 50 classes. The metric on all test sets is Accuracy.

\subsection{Detailed Benchmarks Results}
In our experiments, we evaluated the proposed method across $12$ datasets with both zero-shot \cite{elizalde2024clap2} and state-of-the-art model \cite{deshmukh2024DA} audio domain adaptation (DA). The DA method uses one, five randomly chosen and all unlabelled audio example(s) at test time. As shown in Table \ref{mainTabel}, our method achieves better (+9.38\%) zero-shot classification performance than directly testing the time domain prompt tuning (+5.99\%), with a batch size of 5 (+4.41\%). This illustrates the effectiveness of prompt learning guided by cross-modal consistency as well as contrastive learning regularization.

In challenging datasets such as SESA \citep{SESA} and RAVD \citep{RAVDESS}, the accuracy improved from 73.40\% (DA five) to 76.19\% (ours) and from 34.45\% to 43.71\%, respectively. Additionally, in the commonly used sound classification dataset US8K, the accuracy improved from 85.21\% (DA five) to 86.60\% (ours). This consistent improvement across datasets highlights the effectiveness of our approach. Additionally, the (DA whole) \citep{deshmukh2024DA}, where the prompt was adapted across the entire test set without resetting, achieved an accuracy of 64.57\%, further supporting that our method's superior performance is not due to over-fitting or seeing all test data but rather a result of effective and robust adaptation.

\subsection{Ablation Study}

To further explore the contributions of the two kinds of conditional consistency networks in our method, we conducted an ablation study, separating the networks into without context-aware (c-net) and without domain-aware (d-net) configurations. D-net, when tested individually, achieved an average accuracy of 67.51\%, while c-net achieved 66.96\%. Both networks independently outperform the Zero-Shot \citep{elizalde2024clap2} and (DA one) \citep{deshmukh2024DA} baselines, with notable improvements in datasets like ESC50, where d-net reached 94.55\%, and c-net achieved 94.15\%. However, in some challenging datasets for speech emotion classification task such as CREMA-D, d-net slightly underperformed (32.56\%), compared to 34.13\% for c-net.

When combined (ours), the performance showed substantial gains, achieving 68.83\%, outperforming both individual networks. For example, in SESA, d-net alone achieved 75.24\%, while c-net achieved 73.33\%, but their combination improved to 76.19\%. Similarly, in NSyn, the combined result was 73.54\%, significantly higher than 69.19\% (d-net) and 69.39\% (c-net). This demonstrates that the two networks capture complementary information, leading to stronger test-time adaptation when used together.

\subsection{Cross-domain Generalization}
To verify the effect of the proposed multiple consistency-guided test-time adaptation on the general zero-shot capability of ALMs, we conduct cross-domain ablation experiments. The setup is the same as DA clap, where we train on one target domain and then test its performance on other domains. The results are shown in Table \ref{crossTable}, where each row represents the training domain and each column shows the test results under different datasets. The cross-domain generalization of our method is better than that of DA clap state-of-the-art model under most train/test domain combinations. Besides, our method outperforms zero-shot on average on all training domains. 

\subsection{Conclusion}
We propose an end-end method for multiple consistency-guided test-time adaptation of audio-language models (ALMs), which boosts the performance of classification in comparison with the state-of-the-art methods for $12$ downstream tasks. However, the exploration of unsupervised adaptation on video-audio description and video-audio generation is not well studied. Therefore, we are studying the unsupervised methods for the adaptation of large audio models for these tasks. In order to obtain high-quality audio description and generation, we plan to develop a large-scale video-audio foundation model which is trained with a large-scale dataset.   

\clearpage
{\small 
\bibliographystyle{./IEEEtran}
\bibliography{reference}

\begin{thebibliography}{10}
\providecommand{\url}[1]{#1}
\csname url@samestyle\endcsname
\providecommand{\newblock}{\relax}
\providecommand{\bibinfo}[2]{#2}
\providecommand{\BIBentrySTDinterwordspacing}{\spaceskip=0pt\relax}
\providecommand{\BIBentryALTinterwordstretchfactor}{4}
\providecommand{\BIBentryALTinterwordspacing}{\spaceskip=\fontdimen2\font plus
\BIBentryALTinterwordstretchfactor\fontdimen3\font minus \fontdimen4\font\relax}
\providecommand{\BIBforeignlanguage}[2]{{%
\expandafter\ifx\csname l@#1\endcsname\relax
\typeout{** WARNING: IEEEtran.bst: No hyphenation pattern has been}%
\typeout{** loaded for the language `#1'. Using the pattern for}%
\typeout{** the default language instead.}%
\else
\language=\csname l@#1\endcsname
\fi
#2}}
\providecommand{\BIBdecl}{\relax}
\BIBdecl

\bibitem{elizalde2024clap2}
B.~Elizalde, S.~Deshmukh, and H.~Wang, ``Natural language supervision for general-purpose audio representations,'' \textit{arXiv preprint arXiv:2309.05767}, 2023.

\bibitem{deshmukh2024pengi}
S.~Deshmukh, B.~Elizalde, R.~Singh, and H.~Wang, ``Pengi: An audio language model for audio tasks,'' \textit{arXiv preprint arXiv:2305.11834}, 2023.

\bibitem{gong2024listenthinkunderstand}
Y.~Gong, H.~Luo, A.~H. Liu, L.~Karlinsky, and J.~Glass, ``Listen, think, and understand,'' \textit{arXiv preprint arXiv:2305.10790}, 2023.

\bibitem{wu2024laion}
Y.~Wu, K.~Chen, T.~Zhang, Y.~Hui, M.~Nezhurina, T.~Berg-Kirkpatrick, and S.~Dubnov, ``Large-scale contrastive language-audio pretraining with feature fusion and keyword-to-caption augmentation,'' \textit{arXiv preprint arXiv:2211.06687}, 2022.

\bibitem{clap1}
B.~Elizalde, S.~Deshmukh, M.~A. Ismail, and H.~Wang, ``Clap learning audio concepts from natural language supervision,'' in \emph{ICASSP 2023 - 2023 IEEE International Conference on Acoustics, Speech and Signal Processing (ICASSP)}, 2023, pp. 1--5.

\bibitem{li2023fewshotprompt}
Y.~Li, X.~Wang, and H.~Liu, ``Audio-free prompt tuning for language-audio models,'' \textit{arXiv preprint arXiv:2309.08357}, 2023.

\bibitem{liang2023fewshotadapter}
J.~Liang, X.~Liu, H.~Liu, H.~Phan, E.~Benetos, M.~D. Plumbley, and W.~Wang, ``Adapting language-audio models as few-shot audio learners,'' \textit{arXiv preprint arXiv:2305.17719}, 2023.

\bibitem{coop}
K.~Zhou, J.~Yang, C.~C. Loy, and Z.~Liu, ``Learning to prompt for vision-language models,'' \emph{Int. J. Comput. Vision}, vol. 130, no.~9, p. 2337–2348, sep 2022.

\bibitem{cocoop}
K.~Zhou, J.~Yang, C.~Loy, and Z.~Liu, ``Conditional prompt learning for vision-language models,'' in \emph{2022 IEEE/CVF Conference on Computer Vision and Pattern Recognition (CVPR)}, 2022, pp. 16\,795--16\,804.

\bibitem{promptsrc}
M.~U. Khattak, S.~T. Wasim, M.~Naseer, S.~S. Khan, M.~Yang, and F.~S. Khan, ``Self-regulating prompts: Foundational model adaptation without forgetting,'' \emph{2023 IEEE/CVF International Conference on Computer Vision (ICCV)}, pp. 15\,144--15\,154, 2023.

\bibitem{clip}
A.~Radford, J.~W. Kim, C.~Hallacy, A.~Ramesh, G.~Goh, S.~Agarwal, G.~Sastry, A.~Askell, P.~Mishkin, J.~Clark, G.~Krueger, and I.~Sutskever, ``Learning transferable visual models from natural language supervision,'' in \emph{International Conference on Machine Learning}, 2021.

\bibitem{roy2024consistencyguidedPLVL}
S.~Roy and A.~Etemad, ``Consistency-guided prompt learning for vision-language models,'' \textit{arXiv preprint arXiv:2306.01195}, 2023.

\bibitem{TTT}
Y.~Sun, X.~Wang, Z.~Liu, J.~Miller, A.~A. Efros, and M.~Hardt, ``Test-time training with self-supervision for generalization under distribution shifts,'' in \emph{Proceedings of the 37th International Conference on Machine Learning}, ser. ICML'20.\hskip 1em plus 0.5em minus 0.4em\relax JMLR.org, 2020.

\bibitem{ttt++}
Y.~Liu, P.~Kothari, B.~van Delft, B.~Bellot-Gurlet, T.~Mordan, and A.~Alahi, ``Ttt++: When does self-supervised test-time training fail or thrive?'' in \emph{Advances in Neural Information Processing Systems}, vol.~34.\hskip 1em plus 0.5em minus 0.4em\relax Curran Associates, Inc., 2021, pp. 21\,808--21\,820.

\bibitem{Wang2021TentFT}
D.~Wang, E.~Shelhamer, S.~Liu, B.~A. Olshausen, and T.~Darrell, ``Tent: Fully test-time adaptation by entropy minimization,'' in \emph{International Conference on Learning Representations}, 2021.

\bibitem{TTTmaskae}
Y.~Gandelsman, Y.~Sun, X.~Chen, and A.~A. Efros, ``Test-time training with masked autoencoders,'' in \emph{Proceedings of the 36th International Conference on Neural Information Processing Systems}, ser. NIPS '22.\hskip 1em plus 0.5em minus 0.4em\relax Red Hook, NY, USA: Curran Associates Inc., 2024.

\bibitem{MEMO}
M.~Zhang, S.~Levine, and C.~Finn, ``Memo: Test time robustness via adaptation and augmentation,'' in \emph{Advances in Neural Information Processing Systems}, vol.~35.\hskip 1em plus 0.5em minus 0.4em\relax Curran Associates, Inc., 2022, pp. 38\,629--38\,642.

\bibitem{TPT}
M.~Shu, W.~Nie, D.-A. Huang, Z.~Yu, T.~Goldstein, A.~Anandkumar, and C.~Xiao, ``Test-time prompt tuning for zero-shot generalization in vision-language models,'' in \emph{Advances in Neural Information Processing Systems}, vol.~35.\hskip 1em plus 0.5em minus 0.4em\relax Curran Associates, Inc., 2022, pp. 14\,274--14\,289.

\bibitem{deshmukh2024DA}
S.~Deshmukh, R.~Singh, and B.~Raj, ``Domain adaptation for contrastive audio-language models,'' \textit{arXiv preprint arXiv:2402.09585}, 2024.

\bibitem{Park2019SpecAugmentAS}
D.~S. Park, W.~Chan, Y.~Zhang, C.-C. Chiu, B.~Zoph, E.~D. Cubuk, and Q.~V. Le, ``Specaugment: A simple data augmentation method for automatic speech recognition,'' in \emph{Interspeech}, 2019.

\bibitem{Yamashita_2024_CVPR}
K.~Yamashita and K.~Hotta, ``Mixstyle-based contrastive test-time adaptation: Pathway to domain generalization,'' in \emph{Proceedings of the IEEE/CVF Conference on Computer Vision and Pattern Recognition (CVPR) Workshops}, June 2024, pp. 1029--1037.

\bibitem{HTSAT}
K.~Chen, X.~Du, B.~Zhu, Z.~Ma, T.~Berg-Kirkpatrick, and S.~Dubnov, ``Hts-at: A hierarchical token-semantic audio transformer for sound classification and detection,'' in \emph{ICASSP 2022 - 2022 IEEE International Conference on Acoustics, Speech and Signal Processing (ICASSP)}, 2022, pp. 646--650.

\bibitem{deshmukh2023TACMwithoutAudio}
S.~Deshmukh, B.~Elizalde, D.~Emmanouilidou, B.~Raj, R.~Singh, and H.~Wang, ``Training audio captioning models without audio,'' \textit{arXiv preprint arXiv:2309.07372}, 2023.

\bibitem{Radford2019GPT2}
A.~Radford, J.~Wu, R.~Child, D.~Luan, D.~Amodei, and I.~Sutskever, ``Language models are unsupervised multitask learners,'' 2019.

\bibitem{adamw}
I.~Loshchilov and F.~Hutter, ``Decoupled weight decay regularization,'' in \emph{International Conference on Learning Representations}, 2017.

\bibitem{turian2022hear}
J.~Turian, J.~Shier, H.~R. Khan, B.~Raj, and B.~W. Schuller, ``Hear: Holistic evaluation of audio representations,'' \textit{arXiv preprint arXiv:2203.03022}, 2022.

\bibitem{Piczak2015ESC50}
K.~J. Piczak, ``Esc: Dataset for environmental sound classification,'' \emph{Proceedings of the 23rd ACM international conference on Multimedia}, 2015.

\bibitem{US8K}
J.~Salamon, C.~Jacoby, and J.~P. Bello, ``A dataset and taxonomy for urban sound research,'' in \emph{Proceedings of the 22nd ACM International Conference on Multimedia}, ser. MM '14, New York, NY, USA, 2014, p. 1041–1044.

\bibitem{DCASE2017}
A.~Mesaros, T.~Heittola, A.~Diment, B.~Elizalde, A.~Shah, E.~Vincent, B.~Raj, and T.~Virtanen, ``{DCASE} 2017 challenge setup: tasks, datasets and baseline system,'' in \emph{Proceedings of the Detection and Classification of Acoustic Scenes and Events 2017 Workshop (DCASE2017)}, 2017, pp. 85--92.

\bibitem{Tzanetakis2001GTZAN}
G.~Tzanetakis, ``Automatic musical genre classification of audio signals,'' in \emph{International Society for Music Information Retrieval Conference}, 2001.

\bibitem{CREMAD}
H.~Cao, D.~G. Cooper, M.~K. Keutmann, R.~C. Gur, A.~Nenkova, and R.~Verma, ``Crema-d: Crowd-sourced emotional multimodal actors dataset,'' \emph{IEEE Transactions on Affective Computing}, vol.~5, no.~4, pp. 377--390, 2014.

\bibitem{RAVDESS}
R.~F. Livingstone~SR, ``The ryerson audio-visual database of emotional speech and song (ravdess): A dynamic, multimodal set of facial and vocal expressions in north american english.'' \emph{PLoS One}, 2018.

\bibitem{Gong2022VocalsoundAD}
Y.~Gong, J.~Yu, and J.~R. Glass, ``Vocalsound: A dataset for improving human vocal sounds recognition,'' \emph{ICASSP 2022 - 2022 IEEE International Conference on Acoustics, Speech and Signal Processing (ICASSP)}, pp. 151--155, 2022.

\bibitem{SESA}
T.~Spadini, ``Sound events for surveillance applications,'' 2019.

\bibitem{ns_inst}
J.~Engel, C.~Resnick, A.~Roberts, S.~Dieleman, M.~Norouzi, D.~Eck, and K.~Simonyan, ``Neural audio synthesis of musical notes with wavenet autoencoders,'' in \emph{Proceedings of the 34th International Conference on Machine Learning - Volume 70}, ser. ICML'17.\hskip 1em plus 0.5em minus 0.4em\relax JMLR.org, 2017, p. 1068–1077.

\end{thebibliography}
}
\appendix
To further investigate the performance and limitations of our approach, we provide additional ablation experiments as shown in Table \ref{appendix}.

Firstly, to verify that the effectiveness of combining domain-aware and context-aware prompts is not solely due to the increase in the number of parameters, we conducted an experiment (params *2) in which the neural network width was expanded. We also performed an ablation study on two different loss functions to validate their individual contributions. Finally, we explored the impact of the number of layers in the MLP network on the performance of TTA.

\begin{table*}[ht]
\centering
\caption{Supplemental ablation experiments. w/o means without, and params*2 denotes doubling the number of parameters of the MLP.}
\label{appendix}
\begin{tabular}{l| c | c c c c c c c c c c c c }
\hline
Dataset & Avg.↑ & ESC50 & US8K & D17T4 & SESA & GT MS & Genre & Opera & CREM & RAVD & Vocal & TU17 & NSyn  \\
\hline
Zero-Shot (CLAP )  & 62.93 & 93.90 & 82.30 & 46.60 & 64.95 & 99.20 & 58.40 & 46.60 & 30.00 & 31.54 & 79.97 & 53.80 & 68.00  \\ \hline
DA CLAP (one)  & 64.94 & 93.35 & 85.26 & 50.96 & 73.30 & 99.21 & 61.00 & 47.45 & 29.92 & \textbf{34.25} & 82.14 & 54.19 & 68.28  \\
DA CLAP (five) & 65.92 & 95.05 & 85.21 & 52.30 & 74.35 & 100.0  & 63.20 & 50.42 & 31.27 & 33.19 & 82.40 & 54.38 & 69.23  \\  
DA CLAP (whole)  & 64.57 & 94.15 & 85.83 & 51.33 & 69.52 & 100.0  & 58.06 & 44.49 & \textbf{37.07} & 32.26 & 80.79 & 57.22 & 64.09  \\  \hline
ours(3-layers MLP)   & \textbf{68.83} & 94.65 & 86.60 & 51.69 & 76.19 & 100.0  & 65.87 & 68.64 & 34.29 & 33.77 & 82.21 & 58.46 & 73.54  \\  
-- w/o context-aware        & 67.51 & 94.55 & 84.67 & 50.79 & 75.24 & 100.0  & 61.86 & \textbf{70.34} & 32.56 & 30.06 & 81.37 & \textbf{59.51} & 69.19  \\
   params *2                & 67.33 & 94.70 & 85.15 & 49.76 & 76.19 & 100.0 & 59.06 & 69.07 & 31.77 & 31.69 & 81.09 & 58.89 & 70.63 \\
-- w/o domain-aware        & 66.96 & 94.15 & 82.99 & \textbf{54.29} & 73.33 & 100.0  & 64.67 & 57.63 & 34.13 & 32.91 & \textbf{82.40} & 57.59 & 69.39  \\
   params *2               & 66.84 & 94.15 & 83.02 & 54.11 & 73.33 & 100.0 & 63.96 & 57.63 & 34.05 & 32.91 & 82.29 & 57.04 & 69.60 \\
\hline
-- w/o contrastive loss   & 64.42 & 94.55 & 84.28 & 49.82 & 78.10 & 100.0 & 58.66 & 57.63 & 18.69 & 24.47 & 80.81 & 57.78 & 68.19 \\
-- w/o self-entropy loss  & 67.75 & 94.60 & 86.61 & 50.00 & 78.10 & 100.0 & 58.96 & 68.64 & 32.79 & 31.32 & 80.48 & 58.03 & 73.51 \\
\hline
1-layer MLP               & 68.71 & \textbf{96.00} & \textbf{87.36} & 50.73 & 77.14 & 100.0 & 63.56 & 69.92 & 32.94 & 32.18 & 82.18 & 59.44 & 73.12 \\
2-layers MLP              & 68.50 & 95.35 & 87.04 & 50.54 & 77.14 & 100.0 & 63.96 & 68.64 & 32.81 & 33.36 & 82.23 & 58.09 & 72.80 \\
4-layers MLP              & 68.67 & 94.60 & 86.60 & 53.08 & \textbf{80.95} & 100.0 & \textbf{66.77} & 61.44 & 32.69 & 33.52 & 82.15 & 58.58 & \textbf{73.63} \\
\hline
\hline
\end{tabular}

\end{table*}
\end{document}